\documentclass[prl,aps,amssymb,twocolumn,showpacs,superscriptaddress]{revtex4}
\usepackage{graphicx}

\begin{document}

\title{Comment on ``Experimental Determination of the Statistics of Photons Emitted by a Tunnel Junction"}

\author{B.~Reulet}
\affiliation{Laboratoire de Physique des Solides, CNRS UMR8502, Univ. Paris-Sud 11, F91405 Orsay, France}
\affiliation{D\'{e}partement de Physique, Universit\'{e} de Sherbrooke, Sherbrooke, Qu\'{e}bec, Canada, J1K 2R1}
\author{J.~Gabelli}
\affiliation{Laboratoire de Physique des Solides, CNRS UMR8502, Univ. Paris-Sud 11, F91405 Orsay, France}

\date{\today}

\maketitle
 A very recent article \cite{bad} has addressed a central problem on the statistics of electromagnetic fluctuations generated by a quantum conductor: how does the statistics of electrons crossing a sample influence that of the photons they emit ? It is however not clear that the detection used in \cite{bad} (linear amplifiers followed by square law detectors) is sensitive to emitted photons and not to the overall electromagnetic fluctuations, which include vacuum fluctuations. In \cite{bad} there is an experimental proof that non-symmetrized noise is detected, i.e. emitted photons only. We show that this proof is erroneous: supposing that the detection simply takes the square of the voltage, i.e. is sensitive to vacuum fluctuations, leads to identical results. We demonstrate that all the results of \cite{bad} can indeed be explained in terms of usual gaussian voltage fluctuations instead of photon statistics. Finally, it is found in \cite{bad} that the photon noise is gaussian for a tunnel junction, for which the electron counting statistics is poissonian. We show that this result is beyond the experimental sensitivity by at least one order of magnitude, and thus is also incorrect.

The experimental setup of \cite{bad} is of Hanburry-Brown and Twiss type, and we recall in Fig. 1 a schematics of its canonical version. Such a setup contains a port connected to vacuum (a $50\Omega$ load in the microwave version of the beam splitter, see e.g. \cite{HBT}) which contributes to the signal at the output of the beam splitter.
 If we note $a$, $a^+$ (respectively $b$, $b^+$) the creation and annihilation operators of the photon field in the sample (respectively in the load of the fourth port), the output of the two amplifiers is described by the operators $c_1=G_1(a+ib)+\xi_1$ and $c_2=G_2(a-ib)+\xi_2$ where $G_{1,2}$ are the voltage gains of the amplifiers and $\xi_{1,2}$ the unavoidable noise added by the amplifiers. With the hypothesis of photodetection used in \cite{bad}, the average power after amplification is, omitting the noise added by the amplifiers for simplicity, $P_k\propto G_k\langle n_k\rangle$  with $n_k=c^+_kc_k$. The auto-correlation is $S_{P_kP_k}\propto\langle n_k^2\rangle -\langle n_k\rangle^2$ and the cross-correlation $S_{P_1P_2}\propto\langle n_1n_2\rangle-\langle n_1\rangle \langle n_2\rangle$. Let us now take another point of view and suppose that the amplifiers cannot separate the part of the electromagnetic field that corresponds to real photons from the part that is due to vacuum fluctuations: they simply amplify voltage. Then the average power is $P'_k\propto \frac12 \langle c^+_kc_k+c_kc^+_k\rangle=\langle n_k\rangle+1/2$. The auto-correlation is $S'_{P_kP_k}\propto \langle (n_k+\frac12)^2\rangle -\langle (n_k+\frac12)\rangle^2)$ and the cross-correlation $S'_{P_1P_2}\propto \langle (n_1+\frac12)(n_2+\frac12)\rangle-\langle (n_1+\frac12)\rangle \langle (n_2+\frac12)\rangle$ (note that the operators $n_1$ and $n_2$ commute). Obviously, $S'_{P_kP_k}=S_{P_kP_k}$ and $S'_{P_1P_2}=S_{P_1P_2}$. Thus the assertion in \cite{bad} that the cross-correlations would be different if the symmetrized noise was detected (i.e. if the terms $c_kc^+_k$ could be detected), is incorrect. The mistake of the authors of \cite{bad}, we think, comes from they have not well taken into account the vacuum fluctuations of the $50\Omega$ load (the $b$ term). The dashed curve of their Fig. 4, which does not fit the data, corresponds indeed to $\langle (n_1+\frac12)(n_2+\frac12)\rangle -\langle n_1\rangle \langle n_2\rangle$ and not to $S'_{P_1P_2}$. The origin of the $b$ term in the setup of \cite{bad} is the following (see Fig. 1 of \cite{bad}): the signal coming from the load of the isolator of the upper arm is partially transmitted through the sample and detected in the lower arm, and partially reflected by the sample and detected in the upper arm, and the same occurs for the load of the isolator of the lower arm. These signals contribute to the cross-correlation since they appear in both amplifiers. This is true even if the two loads at are zero temperature and no real photons are emitted.

\begin{figure}
\includegraphics[width= 0.8\columnwidth]{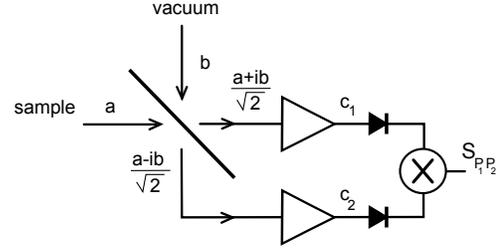}
\caption{Schematics of the conventional Hanburry-Brown and Twiss setup. The vacuum port of the beam splitter is a $50\Omega$ load in microwave circuits.}
\end{figure}

 Now that we have shown that an explanation in terms of voltage fluctuations instead of photons is equally relevant, we will demonstrate that the results of \cite{bad} can be explained by usual voltage noise. Instead of reasoning with photon fields, we simply model the sample and the load by fluctuating voltage sources, respectively noted $v_a$ and $v_b$. The auto-correlation is $S_{P_kP_k}\propto\langle (v_a\pm v_b)^4\rangle-\langle (v_a\pm v_b)^2\rangle^2$ which, for Gaussian variables ($\langle v_{a,b}^4\rangle=3\langle v_{a,b}^2\rangle^2)$ reduces to $S_{P_kP_k}\propto (\langle v_a^2\rangle+\langle v_b^2\rangle)^2$. Let us, as in \cite{bad}, note powers in units of noise temperature, with $T_{eq}$ the noise temperature detected at equilibrium, $T_{eq}=\langle v_a^2\rangle_{eq}+\langle v_b^2\rangle$, and $T=T_{eq}+\Delta T$ the noise temperature detected at finite bias ($v_b$ is not affected by the bias on the sample), i.e. $\Delta T=\langle v_a^2\rangle-\langle v_a^2\rangle_{eq}$. We can rewrite the auto-correlation as $S_{PP}\propto (T_{eq}+\Delta T)^2$, or:
\begin{equation}
\frac{S_{PP}-S_{PP}^{eq}}{S_{PP}^{eq}}=
\left(\frac{\Delta T}{T_{eq}}\right)^2 +2\frac{\Delta T}{T_{eq}}
\end{equation}
This formula coincides with Fig. 3 of \cite{bad}. Concerning the cross-correlation, one obtains $S_{P_1P_2}\propto (\langle v_a^2\rangle - \langle v_b^2\rangle)^2$. This gives formula (5) of \cite{bad} when one reintroduces the individual gains of the amplifiers. Thus all the results of \cite{bad} can be simply explained by gaussian random voltage fluctuations, i.e. one would obtain identical results with e.g. a macroscopic resistor heated by a dc bias.

We now consider the most important result of \cite{bad}: the statistics of photons seems gaussian whereas that of current fluctuations in a tunnel junctions is definitely not. Let us estimate the amplitude of the expected non-Gaussian signal in \cite{bad}. The non-Gaussian aspect of the source appears in the fourth moment of the voltage fluctuations as $\langle v_a^4\rangle=3\langle v_a^2\rangle^2+C_4$ where $C_4$ is the fourth cumulant.
For $eV_{ds}\gg k_BT$ one has $C_4=e^3I\Delta\nu^3$ and $(\langle v_a^2\rangle)^2=(eI\Delta\nu)^2$, with $\Delta\nu=200 MHz$ the bandwidth. The non-Gaussian contribution dominates for $I\lesssim e\Delta\nu$, i.e. a bias $V_{ds}\lesssim16$nV. For a reasonable bias of $10-100\mu$V, the non-Gaussian part of the cross-correlation represents $\sim10^{-4}-10^{-3}$ of the measured signal, and the resolution, of the order of the percent, is clearly not good enough to detect it. Thus the measurement performed in \cite{bad} does not prove that the tunnel junction behaves as a black body. It confirms that any noise, once bandpass-filtered, looks more and more Gaussian as the bandwidth shrinks, by virtue of the central limit theorem.


\end{document}